\documentclass[a4paper,12pt]{article}
\usepackage{amssymb}
\usepackage{mathrsfs}
\usepackage{cite}

\csname @addtoreset\endcsname{equation}{section}

\newcommand{\dd}{\mathrm{d}}

\newcommand{\sder}{\mathcal{D}} 

\newcommand{\dual}{{}^*}

\newcommand{\eq}{\begin{equation}}
\newcommand{\feq}{\end{equation}}
\newcommand{\eqn}{\begin{eqnarray}}
\newcommand{\feqn}{\end{eqnarray}}
\newcommand{\arr}{\begin{eqnarray*}}
\newcommand{\farr}{\end{eqnarray*}}

\def\ep{\epsilon}

\def\si{\sigma}

\begin{document}

\begin{titlepage}
\begin{flushright}
IFUM-777-FT
\end{flushright}
\vspace{.3cm}
\begin{center}
\renewcommand{\thefootnote}{\fnsymbol{footnote}}
{\Large \bf Supersymmetric G\"odel-type Universe in four Dimensions}
\vskip 25mm
{\large \bf {Marco M.~Caldarelli\footnote{marco.caldarelli@mi.infn.it}
and Dietmar Klemm\footnote{dietmar.klemm@mi.infn.it}}}\\
\renewcommand{\thefootnote}{\arabic{footnote}}
\setcounter{footnote}{0}
\vskip 10mm
{\small
Dipartimento di Fisica dell'Universit\`a di Milano\\
and\\
INFN, Sezione di Milano,\\
Via Celoria 16, I-20133 Milano.\\
}
\end{center}
\vspace{2cm}
\begin{center}
{\bf Abstract}
\end{center}
{\small We generalize the classification of all supersymmetric solutions of
  pure $N=2$, $D=4$ gauged supergravity to the case when external sources are
  included.
  It is shown that the source must be an electrically charged dust. We give a
  particular solution to the resulting equations, that describes a G\"odel-type
  universe preserving one quarter of the supersymmetries.
}
\end{titlepage}


In a recent paper \cite{Caldarelli:2003pb}, we classified all supersymmetric
solutions of pure gauged $N=2$ supergravity in four dimensions, generalizing
the results of Tod \cite{Tod:pm} in the ungauged case. Unlike \cite{Tod:pm},
we did not include external sources, and to remedy this will be the main scope
of this letter. The motivation for including sources is twofold. First of all,
it is desirable to have a more complete treatment that considers also this
case.
Second, external dust sources (as well as a negative cosmological constant)
are necessary ingredients to obtain the G\"odel universe \cite{Godel:ga} or its
generalizations \cite{Reboucas:hn}.

These spacetimes, which suffer from closed timelike curves (CTCs), have
become fashionable recently, after the discovery of a
maximally supersymmetric analogue of the G\"odel universe in five dimensions
\cite{Gauntlett:2002nw}.
Generalizations to ten and eleven dimensions have been found in
\cite{Gauntlett:2002fz,Harmark:2003ud}, and a variety of black holes embedded
in G\"odel spacetimes have been discussed in
\cite{Herdeiro,Gimon:2003ms,Brecher:2003wq}.
Various mechanisms of chronology protection 
using holographic screening \cite{Boyda:2002ba}
or supertubes \cite{Drukker} appeared in the literature.
Particle and brane probes in these backgrounds were analyzed in 
\cite{Boyda:2002ba,Drukker,Hikida:2003yd}.
Finally, string propagation on G\"odel universes, which are T-dual to
compactified pp-waves \cite{Boyda:2002ba}, was initiated in \cite{brace}.

The study of these solutions hopefully will shed some light on holography for
spaces more general than AdS, and on the resolution of CTCs
by string theory.
One might thus ask whether there are members of the family of G\"odel-type
solutions in four dimensions \cite{Reboucas:hn} that preserve some
supersymmetry.
As we said, these spacetimes have the common feature of a negative cosmological
constant and external dust sources, so a natural framework to address this
question is $N=2$, $D=4$ gauged supergravity with sources. In ungauged
four-dimensional 
$N=2$ supergravity, supersymmetry imposes the condition that the sources form a
perfect fluid with vanishing pressure \cite{Tod:pm}. If this holds true also
in the gauged case, it is exactly what we want in order to obtain G\"odel-type
solutions. Indeed, it will be shown below that also in the gauged version,
the source must be a charged dust; the only difference is that now the dust
carries only electric charge, whereas the magnetic charge has to vanish.

In what follows, we use the conventions of \cite{Caldarelli:2003pb}. In that
paper, we obtained the general solution of $N=2$, $D=4$ gauged supergravity
that admits at least one Killing spinor $\epsilon$. The solutions fall into two
classes depending on whether the Killing vector
$V^{\mu}=i\bar\epsilon\Gamma^\mu\epsilon$ constructed from
$\epsilon$ is timelike or lightlike. To keep things short, we will generalize
only the timelike case to the inclusion of external sources. The generalization
of the lightlike case will be discussed elsewhere.

Let us briefly recall the results of \cite{Caldarelli:2003pb} for timelike
$V^{\mu}$ (rewritten here in a slightly more compact form). The general
BPS solution reads\footnote{We have chosen the conformal gauge for the
  two-metric $h_{ij}$ appearing in \cite{Caldarelli:2003pb}.}
\begin{eqnarray}
  ds^2 &=& -\frac 4{\ell^2 F\bar F}(dt + \omega_i dx^i)^2
  + \frac{\ell^2 F\bar F}4
  [dz^2 + e^{2\phi}(dx^2 + dy^2)]\,, \label{metric} \\
  {\cal F} &=& \frac{\ell^2}4 F\bar F\left[V\wedge\dd f
    +\dual\left(V\wedge\left(\dd
             g+\frac1\ell \dd z\right)\right)\right]\,, \nonumber
\end{eqnarray}
where $i = 1,2$; $x^1 = x, x^2 = y$, and we defined $\ell F = 2i/(f-ig)$,
with $f=\bar{\epsilon}\epsilon$ and $g=i\bar{\epsilon}\Gamma_5\epsilon$.
Here, $1/\ell$ is the minimal coupling between the graviphoton and the
gravitini, which is related to the cosmological constant by
$\Lambda=-3\ell^{-2}$.
The timelike Killing vector is given by $V = \partial_t$.
The functions $\phi, F, \bar F$, that depend on $x,y,z$, are determined by
the system
\begin{eqnarray}
  \Delta F + e^{2\phi}[F^3 + 3FF' + F''] &=& 0\,, \label{F} \\
  \Delta \phi + \frac12 e^{2\phi}[F'+\bar{F}'+F^2+\bar{F}^2 - F\bar F] &=& 0\,,
  \label{phi}\\
  \partial_z\phi - \mathrm{Re}(F) &=& 0\,,
\end{eqnarray}
where $\Delta = \partial^2_x + \partial^2_y$, and a prime denotes
differentiation with respect to $z$. (\ref{F}) comes from the combined Maxwell
equation and Bianchi identity, whereas (\ref{phi}) results from the
integrability condition for the Killing spinor $\epsilon$. Finally, the shift
vector $\omega$ is obtained from\footnote{It can be shown that the
  integrability conditions for (\ref{omega}) follow from the Maxwell equations,
  even in presence of external sources.}
\begin{eqnarray}
\partial_z \omega_i &=& \frac{\ell^4}8 (F\bar F)^2 \epsilon_{ij}(f\partial_j g
- g\partial_j f)\,, \nonumber \\
\partial_i \omega_j - \partial_j \omega_i &=&
\frac{\ell^4}8 (F\bar F)^2 e^{2\phi}
\epsilon_{ij}\left(f\partial_z g - g\partial_z f + \frac{2f}{\ell}\right)\,,
\label{omega}
\end{eqnarray}
with $\epsilon_{12} = 1$.

If we allow external charged sources carrying electric current $J^e_\mu$ and
magnetic current $J^m_\mu$, the Maxwell equations read
\begin{equation}
\nabla^\mu{\cal F}_{\mu\nu}=-4\pi J^e_\nu\,,\qquad
\nabla_{[\mu}{\cal F}_{\nu\rho]}=\frac{4\pi}3\ep_{\mu\nu\rho}{}^\si J^m_\si\,.
\end{equation}
Note that the inclusion of magnetic currents modifies the Bianchi identity,
so that it is no more possible to define an electromagnetic vector potential
${\cal A}_{\mu}$ in presence of continuous magnetic charge distributions. As
${\cal A}_{\mu}$ appears explicitly in the supercovariant
derivative
\begin{equation}
{\cal D}_{\mu} = \nabla_{\mu} - \frac{i}{\ell}{\cal A}_{\mu} + \frac{1}{2\ell}
            \Gamma_{\mu} + \frac i4 {\cal F}_{\alpha\beta}\Gamma^{\alpha\beta}
            \Gamma_{\mu}
\end{equation}
of gauged supergravity, consistency requires setting the magnetic
current $J^m$ to zero. For the time being, we will keep $J^m$, and show at the
end that conservation of the combined energy-momentum tensor of the
electromagnetic field and the sources also leads to the condition of
vanishing $J^m$. The charged sources carry some energy-momentum
$T^{\rm ext}_{\mu\nu}$. Imposing the integrability conditions, i.e. the
vanishing of the supercurvature \cite{Caldarelli:2003pb},
\begin{eqnarray}
[\sder_{\nu}, \sder_{\mu}]\epsilon &=&
\left[\frac 1\ell(\dual{\mathcal F}_{\nu\mu}\Gamma_5
- i{\mathcal F}_{\nu\mu}) + \frac 1{2\ell^2}\Gamma_{\nu\mu} +
\frac 14{{\mathcal R}^{ab}}_{\nu\mu}\Gamma_{ab}\right. \nonumber \\
&& - {\mathcal F}^{\alpha\beta}{\mathcal F}_{\beta [\nu}\Gamma_{\mu]\alpha}
+ \frac 14 {\mathcal F}_{\alpha\beta}{\mathcal F}^{\alpha\beta}\Gamma_{\nu\mu}
- \frac i{\ell}{{\mathcal F}^{\alpha}}_{[\nu}\Gamma_{\mu]\alpha} \nonumber \\
&& \left.-\frac i2 \Gamma_{\alpha\beta[\nu}\nabla_{\mu]}{\mathcal F}^{\alpha\beta}
- i\nabla_{[\nu}{{\mathcal F}_{\mu]}}^{\alpha}\Gamma_{\alpha}\right]\epsilon = 0\,,
\label{intcond}
\end{eqnarray}
we obtain, after some algebra, the BPS conditions on the currents,
\begin{equation}
J=J^e+iJ^m=\kappa(f-ig)V\,, \label{J}
\end{equation}
where $\kappa$ is up to now an arbitrary real function.
Conservation of $J$ requires that $\kappa$ be time-independent.
The Einstein equations imply that these sources have the dust stress tensor
\begin{equation}
T^{\rm ext}_{\mu\nu}=\kappa V_\mu V_\nu\,.
\end{equation}
As the Einstein tensor is conserved, the same must be true for the
combined energy-momentum tensor of the electromagnetic field and the
sources. This leads to $g=0$ and thus, from (\ref{J}), to
$J^m=0$ \footnote{The violation of the divergence
of the stress tensor is proportional to $\ell^{-1}$, and thus vanishes
in the ungauged case $\ell \to \infty$, where $J^m = 0$ is no more necessary
\cite{Tod:pm}.},
which, as pointed out above, is already necessary in order to define a gauge
potential ${\cal A}$. We set thus $J^m = 0$ in what follows.

We can now follow closely the discussion of the sourceless timelike
solutions in \cite{Caldarelli:2003pb}. In fact, the only difference is in
Maxwell's equations which include now the sources. The general supersymmetric
solution with timelike Killing vector $V^\mu$ is therefore still given by
(\ref{metric}), (\ref{phi}) and (\ref{omega}), but with (\ref{F}) replaced by
\begin{equation}
\Delta F + e^{2\phi}[F^3 + 3FF' + F'' + 4\pi\kappa F] = 0\,.
\end{equation}
Furthermore, the condition $g=0$ yields $\bar F = -F$. The imaginary part of
equation (\ref{phi}) implies then $F' = 0$, so that
the functions $F$ and $\phi$ are determined by solving the system
\begin{eqnarray}
\Delta F + e^{2\phi}[F^3 + 4\pi\kappa F] &=& 0\,, \label{Fmod} \\
\Delta \phi + \frac 32 e^{2\phi}F^2 &=& 0\,. \label{phimod}
\end{eqnarray}

Let us assume that $\kappa$ is a positive constant. Then a simple solution
to (\ref{Fmod}) is given by $F=i\sqrt{4\pi\kappa}$. 
Equation (\ref{phimod}) becomes the Liouville equation and describes a
two-manifold $h_{ij}=e^{2\phi}\delta_{ij}$ of constant curvature
$-12\pi\kappa$.
For convenience, we choose the parabolic Liouville solution to cast
the transverse metric in Poincar\'e coordinates.
This leads to the metric
\begin{equation}
ds^2=-\frac1{\pi\kappa\ell^2}
          \left(dt-\frac{2\ell^2\sqrt{\pi\kappa}}{3x}\ dy\right)^2
          +\pi\kappa\ell^2\ dz^2
          + \frac{\ell^2}{6x^2}\left(dx^2 + dy^2\right)\,, \label{metric2}
\end{equation}
with uniform magnetic flux through the transverse manifold,
\begin{equation}
{\cal F} = \frac{\ell}{6x^2}\ dx\wedge dy\,.
\end{equation}
A detailed analysis shows that this solution preserves exactly $1/4$ of the
original supersymmetry.
The resulting spacetime has the G\"odel-type metric with
vorticity $\Omega=2/\ell$ and parameter $m=\sqrt6/\ell$ given in
\cite{Reboucas:hn}, and represents a  spacetime homogeneous universe with
rigidly rotating dust and a magnetic field.
The presence of CTCs is most obvious in cylindrical coordinates,
\begin{equation}
 ds^2=-\left(dt-\frac{4\Omega}{m^2}\sinh^2\left(\frac{mr}2\right)
 d\varphi\right)^2+\frac1{m^2}\sinh^2(mr)\ d\varphi^2+dr^2+dz^2\,.
\end{equation}
For sufficiently large $r$ the vector $\partial_\varphi$ is timelike, and
its integral curves become CTCs.

In the ungauged case, the charged dust source can be interpreted as the dilaton
field of the $N=4$ theory \cite{Kallosh:1992ii}. It would be interesting to see
whether such an identification is possible also in the gauged case, for example
by embedding the solution into $N=4$ gauged supergravity.


\section*{Acknowledgements}
\small

This work was partially supported by INFN, MURST and
by the European Commission RTN program
HPRN-CT-2000-00131, in which M.~M.~C.~and D.~K.~are
associated to the University of Torino.
\normalsize



\begin{thebibliography}{99}

\bibitem{Caldarelli:2003pb}
M.~M.~Caldarelli and D.~Klemm,
JHEP {\bf 0309} (2003) 019.

\bibitem{Tod:pm}
K.~P.~Tod,
Phys.\ Lett.\ B {\bf 121} (1983) 241.

\bibitem{Godel:ga}
K.~G\"odel,
Rev.\ Mod.\ Phys.\  {\bf 21} (1949) 447.

\bibitem{Reboucas:hn}
M.~J.~Rebou\c{c}as and J.~Tiomno,
Phys.\ Rev.\ D {\bf 28} (1983) 1251.

\bibitem{Gauntlett:2002nw}
J.~P.~Gauntlett, J.~B.~Gutowski, C.~M.~Hull, S.~Pakis and H.~S.~Reall,
hep-th/0209114.

\bibitem{Gauntlett:2002fz}
J.~P.~Gauntlett and S.~Pakis,
JHEP {\bf 0304} (2003) 039.

\bibitem{Harmark:2003ud}
T.~Harmark and T.~Takayanagi,
Nucl.\ Phys.\ B {\bf 662} (2003) 3.

\bibitem{Herdeiro}
C.~A.~Herdeiro,
Nucl.\ Phys.\ B {\bf 665} (2003) 189;
C.~A.~Herdeiro, hep-th/0307194.

\bibitem{Gimon:2003ms}
E.~G.~Gimon and A.~Hashimoto,
Phys.\ Rev.\ Lett.\  {\bf 91} (2003) 021601.

\bibitem{Brecher:2003wq}
D.~Brecher, U.~H.~Danielsson, J.~P.~Gregory and M.~E.~Olsson,
arXiv:hep-th/0309058.

\bibitem{Boyda:2002ba}
E.~K.~Boyda, S.~Ganguli, P.~Ho$\rm\check{r}$ava and U.~Varadarajan,
Phys.\ Rev.\ D {\bf 67} (2003) 106003.

\bibitem{Drukker}
N.~Drukker, B.~Fiol and J.~Simon,
hep-th/0306057;
N.~Drukker, B.~Fiol and J.~Simon, hep-th/0309199.

\bibitem{Hikida:2003yd}
Y.~Hikida and S.~J.~Rey,
Nucl.\ Phys.\ B {\bf 669} (2003) 57.

\bibitem{brace}
D.~Brace, C.~A.~Herdeiro and S.~Hirano,
hep-th/0307265.

\bibitem{Kallosh:1992ii}
R.~Kallosh, A.~D.~Linde, T.~Ort\'{\i}n, A.~Peet and A.~Van Proeyen,
Phys.\ Rev.\ D {\bf 46} (1992) 5278.

\end{thebibliography}
\end{document}